\documentclass[aps,prl,twocolumn,showpacs,superscriptaddress]{revtex4}

\usepackage{graphicx}

\begin{document}

\title{Spin-orbit interaction in chiral carbon nanotubes probed in pulsed magnetic fields}

\author{S. H. Jhang}
\affiliation{Institute of Experimental and Applied Physics, University of Regensburg, 93040
Regensburg, Germany}

\author{M. Marganska}
\affiliation{Institute for Theoretical Physics, University of Regensburg, 93040 Regensburg,
Germany}

\author{Y. Skourski}
\affiliation{Dresden High Magnetic Field Laboratory, Forschungszentrum Dresden-Rossendorf, 01314
Dresden, Germany}

\author{D. Preusche}
\affiliation{Institute of Experimental and Applied Physics, University of Regensburg, 93040
Regensburg, Germany}

\author{B. Witkamp}
\affiliation{Kavli Institute of Nanoscience, Delft University of Technology, 2628 CJ Delft, The
Netherlands}

\author{M. Grifoni}
\affiliation{Institute for Theoretical Physics, University of Regensburg, 93040 Regensburg,
Germany}

\author{H. van der Zant}
\affiliation{Kavli Institute of Nanoscience, Delft University of Technology, 2628 CJ Delft, The
Netherlands}

\author{J. Wosnitza}
\affiliation{Dresden High Magnetic Field Laboratory, Forschungszentrum Dresden-Rossendorf, 01314
Dresden, Germany}

\author{C. Strunk} \altaffiliation{e-mail:christoph.strunk@physik.uni-regensburg.de}
\affiliation{Institute of Experimental and Applied Physics, University of Regensburg, 93040
Regensburg, Germany}

\begin{abstract}
The magneto-conductance of an open carbon nanotube (CNT)-quantum wire was measured in pulsed
magnetic fields. At low temperatures we find a peculiar split magneto-conductance peak close to the
charge neutrality point. Our analysis of the data reveals that this splitting is intimately
connected to the spin-orbit interaction and the tube chirality. Band structure calculations suggest
that the current in the peak regions is highly spin-polarized, which calls for application in
future CNT-based spintronic devices.
\end{abstract}

\pacs{73.63.Fg, 75.47.-m, 73.23.Ad, 85.75.-d}

\maketitle

A source of spin-polarized electrons is one of the important building blocks of a future spin-based
electronics \cite{Zutic}. Very high degrees of polarization can potentially be achieved by
exploiting spin-orbit interaction (SOI) \cite{Kato,Wunderlich}. Based on the low atomic number $Z =
6$ of carbon the spin-orbit interaction in carbon nanotubes (CNTs) was mostly believed to be very
weak, until a recent experiment \cite{Kuemmeth} has demonstrated the effect of spin-orbit
interaction in clean CNT quantum dots. Evidence for the spin-orbit splitting in simple
magneto-conductance (MC) measurements has not yet been reported.

In this Letter, we present MC data for the complementary situation
of an \textit{open} CNT-quantum wire obtained in pulsed magnetic
fields. In a parallel magnetic field $B_\parallel$, a small band-gap CNT evolves
via a metallic state into a semiconducting one, resulting in a
typical peak in the MC \cite{Fedorov2,Nakanishi}. In one of our
tubes, however, we observed a splitting of this MC-peak into two
peaks at low temperature. Recording MC-traces at different $V_g$
shows that the splitting vanishes when moving away from the charge neutrality
point (CNP). A thorough comparison to band structure calculations
reveals that the splitting is explained by the SOI, which becomes
strong for small tube diameters. An interesting implication of our
analysis is the prediction of a highly spin-polarized current in the peak
regions.

The experiments have been performed on devices made of individual
CNTs prepared on Si/SiO$_{2}$/Si$_{3}$N$_{4}$ substrates. The
heavily p-doped Si was used as a back gate and the thickness of the
insulating layer was 350 nm. CNTs were grown by means of a chemical
vapor deposition method \cite{Kong} and Pd (50 nm) electrodes were
defined on top of the tubes by e-beam lithography. In order to
exclude strain effects on the band structure \cite{Minot2}, only straight and long
($\sim$ 50 $\mu$m) CNTs were selected for devices and the distance
between two Pd electrodes was $\sim$ 500 nm. The $dc$
magneto-conductance was studied in pulsed magnetic fields
of up to 60 T, applied parallel to the tube axis. The accuracy of
the alignment was $\sim \pm 5^\circ$ (See EPAPS for further
experimental details).

\begin{figure}[t]
\includegraphics[width=8.6cm]{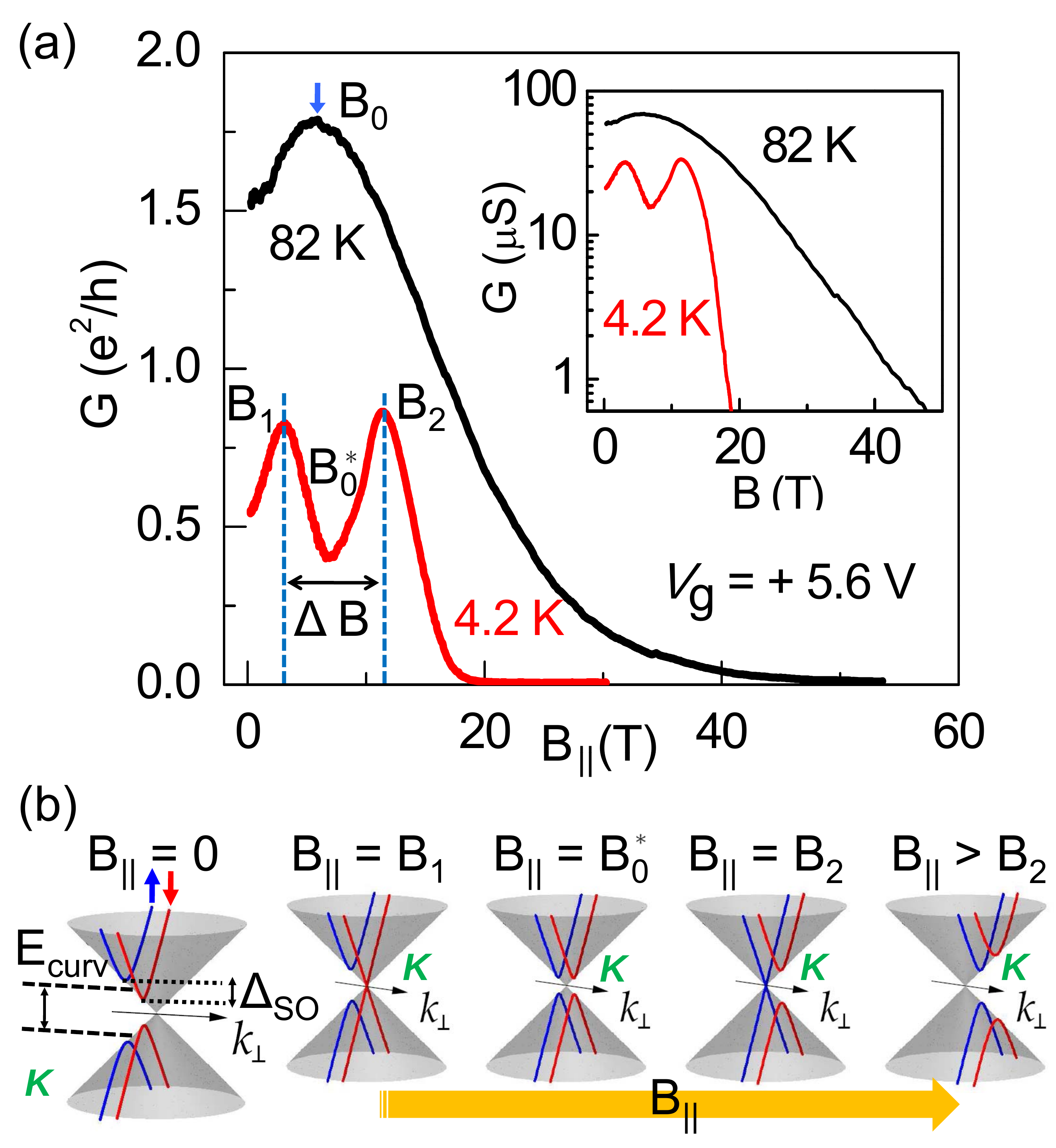}
\caption{ (a) Magneto-conductance (MC) of a small-bandgap CNT device near the charge neutrality
point (CNP) measured at 82~K and 4.2~K. The inset shows the MC in a semi-log scale. The observed
double-peak in $G(B_\parallel)$ defines the characteristic fields $B_0^\ast, B_1$ and $B_2$. (b)
One of the Dirac cones near the $K$-points, intersected by lines of allowed $k_\perp$ values for a
small-bandgap CNT with spin-orbit interaction (SOI). Spin-up (blue) and spin-down (red) bands
display a curvature induced band gap $E_\mathrm{curv}$ and are
 separated by $\Delta_\mathrm{SO}$, due to the SOI.
 With increasing $B_\parallel$, spin-split sub-bands shift due to the AB effect and cross
the $K (K')$ point, successively closing the energy gap at $B_{\mathrm{1}}$ and $B_\mathrm{2}$.
For simplicity, the Zeeman term and $\varepsilon_{\text{SO}}$, an additional Zeeman-like term induced by the SOI,
are neglected.
However, $B_1$ and $B_2$ are not affected by those terms (see EPAPS).
}\label{fig:Tdep}
\end{figure}

Figure~\ref{fig:Tdep}a shows the magneto-conductance $G(B_\parallel)$
of a small-bandgap CNT device located near the
CNP (diameter $d \sim$ 1.5~nm). At 82 K, the conductance $G$ of the tube initially increases to
reach a maximum at $B_{\text{0}}$ = 5.9~T, then it exponentially
drops to zero at higher fields due to the Aharonov-Bohm (AB) effect
\cite{Ajiki,Fedorov2}. Interestingly, when the device was cooled
down to 4.2~K, the conductance maximum $G_{\text{max}}$ at
$B_{\text{0}}$ was split into two distinct peaks at magnetic fields
$B_{\text{1}}$ = 3.1~T and $B_{\text{2}}$ = 11.1~T. We note that
these peaks are symmetric with respect to the conductance dip at
$B_{\text{0}}^\ast$ $\approx$ 7~T, and $G$($B_{\text{0}}^\ast$) is
similar in magnitude to $G$($B_\parallel$ = 0). The key to the
explanation of the data lies in the magnetic field dependence of the
one-dimensional band structure.

A specific CNT is uniquely labeled by the chiral indices $(n,m)$, which define the chiral angle
$\theta$ and the quantized values of the transversal wavevector $k_\perp$ \cite{Charlier}. The
values of $k_\perp$, combined with the graphene dispersion cones, determine the
quasi-onedimensional band structure of the CNTs. A given CNT is metallic if the lines of allowed
$k_\perp$ cross the Dirac points $K,K'$; otherwise it is semiconducting.  For nominally metallic CNTs ($n-m = 3l$, with $l$ an integer),
the dispersion relation $E(k_\parallel)$ near the Dirac points reads
\cite{Kuemmeth,Ando,Bulaev,Izumida,Jeong}:
\begin{eqnarray}\nonumber
\label{eq:cone-complete}
 E(k_\parallel)& = & \pm \hbar v_F \sqrt{ k_\parallel^2 + k_\perp^2 }
+ \left(\frac{g}{2} \mu_B B_\parallel + \tau\, \varepsilon_{\text{SO}}\right) \sigma, \;\\
k_\perp & = & k_{\text{AB}} + k_\perp^0 + k_{\text{SO}}\;,
\end{eqnarray}
where $k_\parallel$ is the wave vector parallel to the tube axis, $v_F$
the Fermi velocity, $\frac{g}{2} \mu_B  B_\parallel \sigma$
being the Zeeman term with $\sigma=\pm1$ for spin
parallel/antiparallel to the tube axis, and $\tau = \pm 1$ for the
$K$ and $K'$ Dirac points. The transversal wave vector $k_\perp$
contains three distinct contributions, which are discussed below.

The Aharonov-Bohm flux $\phi_{\text{AB}} =  B_\parallel \pi
d^2/4$ results in a shift $k_{\text{AB}} = (2/d) (\phi_{\text{AB}} /
\phi_0)$ of $k_\perp$, where $\phi_0=h/e$ is the flux quantum.
 Therefore, one can convert a metallic CNT into a
semiconducting one, or vice versa, by tuning the allowed values of $k_\perp$ with a magnetic field
parallel to the tube axis \cite{Ajiki,Coskun,Zaric,Lassagne,Fedorov2}.

In addition, curvature \cite{Kane} affects the
allowed values of $k_\perp$ and induces small band gaps in nominally metallic CNTs. The
curvature-induced shift \cite{Kane,Kleiner}
$k_\perp^0 = - \, \tau\, a_{\text{0}}\, \cos(3\theta)/(2d)^{2}$
of the allowed $k$-states results in a band gap
$E_{\text{curv}}=2\hbar v_F \vert k_\perp^0 \vert$ at
$B_\parallel=0$, where $a_{\text{0}}$ is the C$-$C bond length. 

A second consequence of curvature is a spin-dependent shift
\begin{equation}\label{eq:k_SO}
k_{\text{SO}} = - \, \sigma \,(2/d) \,(\phi_{\text{SO}}/{\phi_0}),
\end{equation}
of $k_\perp$ by the spin orbit interaction \cite{Kuemmeth,Ando,Bulaev,Izumida,Jeong,Huertas,Chico2} which removes  the four-fold spin and $K,K'$-degeneracy in favor of two Kramers doublets
corresponding to parallel and antiparallel alignment of orbital and
spin magnetic moments.
This SOI-induced shift in $k_\perp$ is equivalent to the presence of an AB flux $\phi_{\text{SO}}\approx 10^{-3} \phi_0$ \cite{Kuemmeth,Ando},
and produces a spin-orbit
energy splitting $\Delta_{\text{SO}}=2\,\hbar v_F \,|k_{\text{SO}}|$.
For a CNT with $d \sim 1\;$nm, $\phi_{\text{SO}}$ corresponds to $\simeq 5\,$T,
while $\phi_0$ is $\simeq 5000\,$T.

In contrast, the term with $\varepsilon_{\text{SO}}= - \, \delta \,
\cos(3\theta)/d$, added to the root in the
Eq.~\ref{eq:cone-complete} (like the Zeeman term), solely shifts the
energy but not $k_\perp$, leading to an asymmetric spin-orbit
energy splitting for the hole
($\Delta_{\text{SO}}+2\varepsilon_{\text{SO}}$) and the electron
band ($\Delta_{\text{SO}}-2\varepsilon_{\text{SO}}$) of chiral
metallic tubes \cite{Izumida,Jeong}.
As $\varepsilon_{\text{SO}}$
contains the factor cos(3$\theta$), it is small for near armchair
tubes. The parameter $\delta$ ranges from 0.3-0.7~nm~meV
\cite{Izumida,Jeong}.

\begin{figure*}
\includegraphics[width=17.5cm]{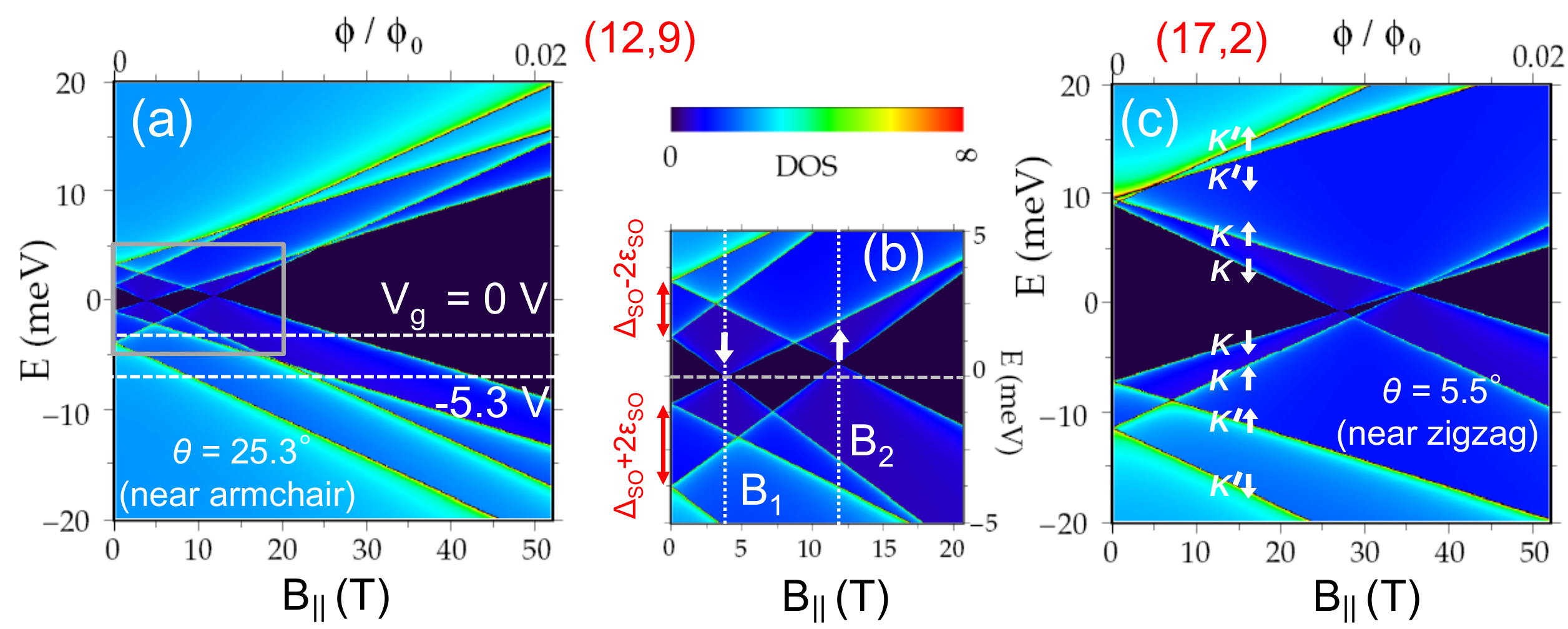}
\caption{ (a) Calculated density of states in a parallel magnetic field for the (12,9) CNT. (b)
Zoom into the area bounded by the gray box clearly shows that the band gap is closed at
$B_{\text{1}}$ and $B_{\text{2}}$ in good agreement with the peak positions observed in
Fig~\ref{fig:Tdep}a. White arrows indicate the spin-polarization of the bands near the crossing
points $B_1$ and $B_2$. (c) DOS calculated for the (17,2) CNT. Due to the larger curvature-induced
band gap of the (17,2) tube with $\theta$ close to 0$^\circ $, the Zeeman-energy splitting at
$B_{\text{0}}$ ($g \mu_B B_{\text{0}}\approx 3.7\;$meV) is larger than the spin-orbit energy
splitting ($\Delta_{\text{SO}}\approx2.33\;$meV). Hence, the SOI-induced peak splitting is
pronounced for the CNTs with chiral angles close to 30$^\circ$. Orbital ($K$ and $K'$) and spin
states (white arrows) are indicated. }\label{fig:DOS}
\end{figure*}

The resulting evolution of the band structure in magnetic field is
visualized in Fig.~\ref{fig:Tdep}b. At zero-field the band gap
$E_{\text{g}}^0=E_{\text{curv}} - \Delta_{\text{SO}}$ is reduced by
the SOI. With the application of $B_\parallel$, the two spin
sub-bands separated by the SOI cross the corner point of the
Brillouin zone (either at $K$ or $K'$), thus explaining two
subsequent MC-peaks at $B_{\text{1}}$ and $B_{\text{2}}$.
In between, a conductance dip appears at $B_{\text{0}}^\ast$ when
the spin sub-bands are located symmetrically around the corner
point. If the Zeeman-like terms in Eq.~\ref{eq:cone-complete} are
neglected the energy gap has a local maximum  at $ k_{\text{AB}} =
-k_\perp^0$ corresponding to $E_{g}(B_{\text{0}}^\ast) \approx
\Delta_{\text{SO}}$. The distance between the two peaks, $\Delta B =
(4/\pi d^2) \, \Delta \phi_{\text{AB}}$, is determined by $\Delta
\phi_{\text{AB}}$ = 2 $\phi_{\text{SO}}$ (the factor 2 comes from
$\sigma = \pm 1$). For the observed values of $\Delta B$ = 8~T,
and $d=1.5\;$nm we find
\begin{equation}
\label{eq:phi_SO} \phi_{\text{SO}} = \frac{\pi d^{2} \Delta B}{8} \approx 1.7 \, \times 10^{-3}
\phi_0. \;
\end{equation}
For a conservative confidence interval of $\pm0.5\;$nm for $d$ determined with an atomic force
microscope one obtains $0.76< 10^3\phi_{\text{SO}}/\phi_0<3$ compatible with previous studies
\cite{Ando,Kuemmeth,Huertas}. Eqs.~(\ref{eq:cone-complete}), (\ref{eq:k_SO}) and
(\ref{eq:phi_SO}) result in the energy splitting $\Delta_{\text{SO}}$ at $B_\parallel=0$
(assuming $k_\parallel=0$)
\begin{equation}
\label{eq:Delta_SO} \Delta_{\text{SO}}  = \frac{4\hbar v_F}{d}
\frac{\phi_{\text{SO}}}{\phi_0}\approx 2.5 \pm 0.8 \,\text{meV}. \;
\end{equation}
This value corresponds to $\sim$ 30 K and explains the disappearance
of the double-peak structure and the single conductance maximum at
$B_\parallel=B_0\simeq B^\ast_0$ for the 82~K trace of
Fig.~\ref{fig:Tdep}. Because $\Delta_{\text{SO}}$ is inversely
proportional to the diameter, it becomes large for small-diameter
tubes \cite{SO}.

With further increase of $\phi_{\text{AB}}$ the energy gap $E_{g}$
linearly opens again as both orbital sub-bands gradually move away
from the corner points of the Brillouin zone. The exponential
decrease of $G$ at high fields (the inset of Fig.~\ref{fig:Tdep}a)
is thus explained by charge carriers thermally activated over the
magnetic-field-induced band gap, as described by previous authors
\cite{Minot,Fedorov2}. Since the inset in Fig.~\ref{fig:Tdep}a
suggests that the conductance depends exponentially on the band gaps
$E_g^0$ and $E_g(B_0^\ast)$, they dominate the conductance at
$B_\parallel=0$ and $B_\parallel=B_0^\ast$. Hence, the approximate
equality $G(0)\approx G(B_0^\ast)$ inferred from
Fig.~\ref{fig:Tdep}a suggests the following relation between the two
energy scales $E_{\text{curv}}$ and $\Delta_\mathrm{SO}$:
\begin{equation}
\label{eq:E_G_0} \frac{E_{g}^0}{E_{g}(B_{\text{0}}^\ast)} = \frac{E_{\text{curv}}
-\Delta_{\text{SO}}} {\Delta_{\text{SO}}} \approx  1. \;
\end{equation}

We now turn to the discussion of the effect of tube chirality. The
strong dependence of $B_{\text{0}}$ and $B^\ast_{\text{0}}$ on the
chirality can be used to identify the chiral indices of
small-bandgap CNTs \cite{Fedorov2}. Out of 53 small-bandgap CNTs
with $d = 1.5 \pm 0.5\;$nm, only five tubes [(12,9), (13,10),
(16,10), (18,9) and (19,10)] display values of $B_{\text{0}}\approx
5$-8~T compatible with our data, while $B_{\text{0}}$ can take much
larger values for other CNTs, e.g., the (12,3) and (17,2) tubes.

Table~I lists values of $\phi_{\text{SO}}$ and $\Delta_{\text{SO}}$
for these chiralities, calculated from Eqs.~(\ref{eq:phi_SO}) and
(\ref{eq:Delta_SO}) and the observed $\Delta B$ = 8~T. The
$\phi_{\text{SO}}$ of the (12,9) tube is closest to
$\phi_{\text{SO}}\approx 10^{-3} \phi_0$, predicted in
Ref.~\onlinecite{Ando} and measured in Ref.~\onlinecite{Kuemmeth}.
When we further take into account the condition $E_{\text{curv}}
\approx2 \, \Delta_{\text{SO}}$ for the CNT measured
(Eq.~(\ref{eq:E_G_0})), we realize that the (12,9) and (18,9) tubes
satisfy this constraint best.

Taking the (12,9) tube with the chiral angle $\theta=25.3^\circ$ (close to the armchair
configuration) as the most probable candidate, we calculated the density of states (DOS) in a
parallel magnetic field.
For comparison, we show the DOS of a
(17,2) tube, which has almost the same diameter but a very different chiral angle
$\theta=5.5^\circ$ close to the zigzag-configuration.

From Fig.~\ref{fig:DOS}, it becomes apparent that in an applied magnetic field the band edges change with four distinct slopes
away from the two Kramers doublets both in the electron and hole bands, reflecting the orbital
and Zeeman splitting. The DOS calculated for the (12,9) tube explains the evolution of the
magneto-conductance very well. The band gap is closed at
$B_{\text{1}}$ by the spin-down and subsequently at $B_{\text{2}}$ by the spin-up sub-band, in very
good agreement with the observed double-peaks at the CNP. The calculated
energy gaps at zero-field and at $B_{\text{0}}^\ast$ agree with Eqs.~(\ref{eq:Delta_SO}) and
(\ref{eq:E_G_0}).
\begin{table}[b]
\caption{\label{tab:table1} Determination of the chirality from $G(B_\parallel)$ at the charge
neutrality point. $\phi_{\text{SO}}$ was calculated from Eq.~(\ref{eq:phi_SO}) with the measured
value of $\Delta B = 8\;$T and used as input for the evaluation of $\Delta_{\text{SO}}$ using
Eq.~(\ref{eq:Delta_SO}). }
\begin{ruledtabular}
\begin{tabular}{ccccccc}
(n,m) & $d$ & $\theta$ & $E_{\text{curv}}$ &
$B_{\text{0}},B^\ast_{\text{0}}$ & $\phi_{\text{SO}}$
&$\Delta_{\text{SO}}$
\\  & ( nm ) & ( $^\circ $ ) & ( meV ) & ( T ) & ( $10^{-3}\phi_0$ ) & ( meV )\\
\hline
(12,9) & 1.43 & 25.3 & 4.6  & 7.8  & 1.54 & 2.36\\
(13,10) & 1.56 & 25.7 & 3.5 & 5.5  & 1.84 & 2.58\\
(16,10) & 1.78 & 22.4 & 4.7 & 6.4 & 2.39& 2.94\\
(18,9) & 1.86 & 19.1 & 6.0 & 7.8 & 2.63 & 3.08\\
(19,10) & 2.00 & 19.8 & 4.9 & 5.9 & 3.02 & 3.30\\
\hline
(12,3) & 1.08 & 10.9 & 28.1 & 63.1 & 0.88 & 1.78 \\
(17,2) & 1.42 & 5.5 & 18.5 & 31.6 & 1.51 & 2.33\\
\end{tabular}
\end{ruledtabular}
\end{table}
On the other hand, the DOS calculated for the (17,2) CNT predicts a
significantly reduced intermediate gap region in
Fig.~\ref{fig:DOS}c, in spite of almost the same $d$ and
$\Delta_{\text{SO}}$, when compared with the (12,9) tube. The (17,2)
tube has a much larger curvature-induced gap
$E_{\text{curv}}\approx18.5$~meV, resulting in a much higher
$B_{\text{0}}\approx31.6$~T. Because the spin-orbit gap competes
with the Zeeman splitting, the peak splitting in $G(B_\parallel)$ is
most pronounced for near armchair tubes with chiral angles close to 30$^\circ$.
\begin{figure}
\includegraphics[width=8.6cm]{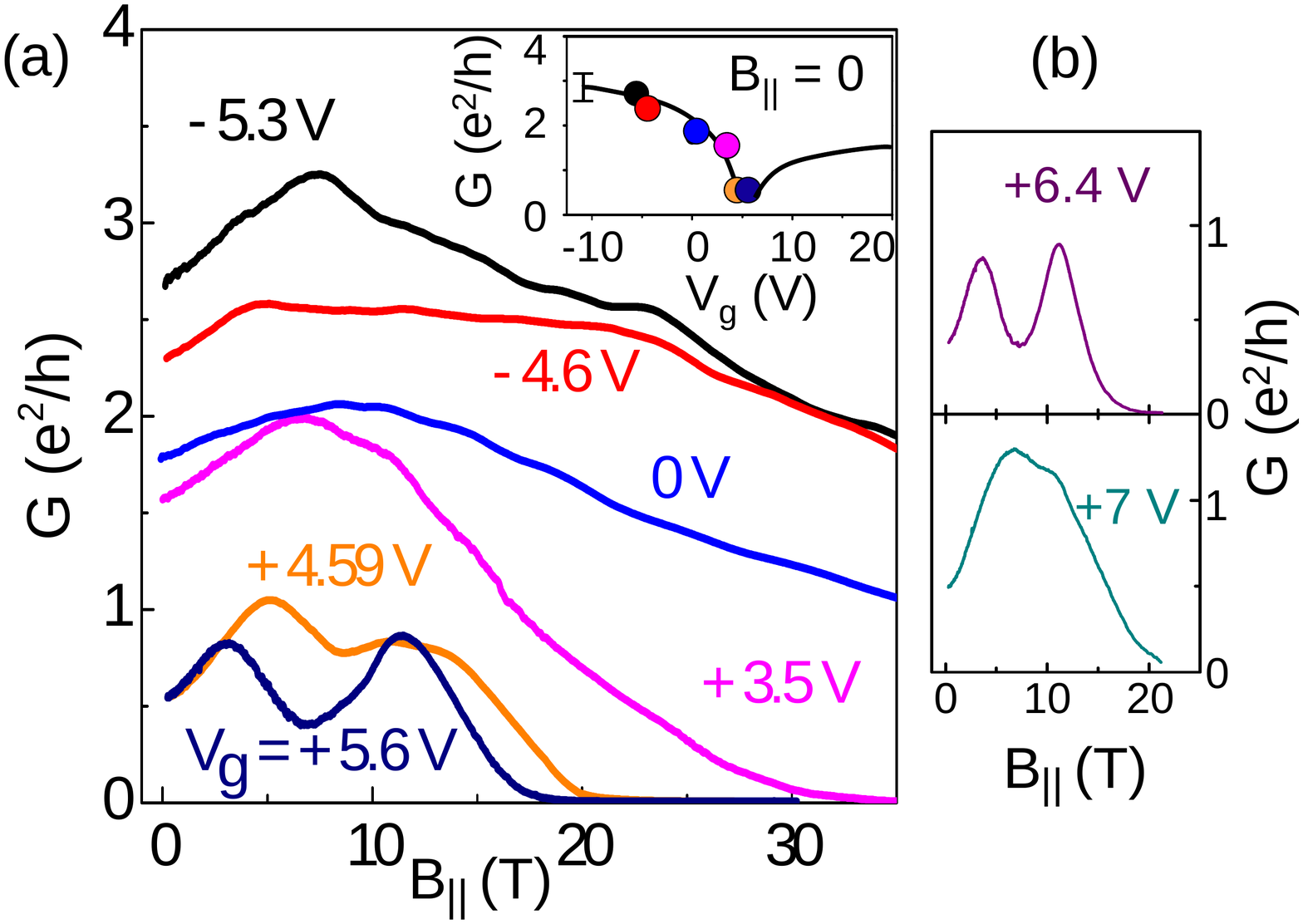}
\caption{ (a) $G(B_\parallel)$ traces at 4.2 K for the hole side of
the CNP. As the  band gap grows with $B_\parallel$ at high fields,
the gate characteristic $G(V_{g})$ exhibits the behavior of a p-type
CNT field-effect-transistor with on-off conductance ratio of several
orders of magnitude. The insert shows $G(V_{g})$ at $B_\parallel$ =
0. Each colored dot corresponds to the same colored trace of
$G(B_\parallel)$. The black solid line in the insert is a fit to
experimental points (see error bar) given as a guide line to the
eye. (b) $G(B_\parallel)$ curves for the electron side of the CNP.
The double-peak structure is observed only when the Fermi energy is
tuned close to the CNP at $V_{g}$ $\simeq +6\;$V.
}\label{fig:Gate_dep}
\end{figure}

So far, all our discussion focused on the vicinity of the charge neutrality point. As a crucial test of our analysis,
we traced the evolution of the $G(B_\parallel)$ curves for various values of gate voltage ($V_{g}$).
In Fig.~\ref{fig:Gate_dep}a, magneto-conductance traces at 4.2~K are displayed for the hole side of
the CNP. Deeply inside the hole band the conductance exceeds 3$e^{2}/h$, close to the theoretical
limit of 4$e^{2}/h$. This shows that our device is in the ballistic regime, where the conductance
is determined by the number of available sub-bands with an average transmission probability of $\sim
0.8$. At $B_\parallel$ = 0, the hole conductance is around 3$e^{2}/h$ and diminishes down to $\sim
0.5 e^{2}/h$ as $E_{\text{F}}$ is tuned towards the CNP ($V_{g}^*$ $\sim$ +6~V). While the
magneto-conductance is initially positive at low fields, it becomes negative at high fields
($B_\parallel \gg B_{\text{0}}$) for all gate voltages, indicating the growth of $E_{g}$ due to the
AB effect. At $B_\parallel >$ 30~T, the gate
characteristic $G(V_{g})$ exhibits the behavior of a p-type CNT field-effect-transistor with an on-off
conductance ratio $> 10^3$. The double-peak structure is pronounced only in
the vicinity of the CNP (+5.6~V $\leq$ $V_{g}^*$ $\leq$ +6.4~V). Two additional $G(B_\parallel)$
curves, presented in Fig.~\ref{fig:Gate_dep}b, show that the two peaks merge again into one
as $E_{\text{F}}$ is shifted to the electron side across the CNP.

Approximating the backgate coupling to the Fermi energy shift as $\Delta E_{\text{F}}$ $\approx 0.7
\times 10^{-3}\Delta V_{g}$, the DOS calculation matches the gate voltages in
Fig.~\ref{fig:Gate_dep}. For instance, the conductance kinks observed at 8 and 24~T for the
$G(B_\parallel)$ curve at $V_{g}$= -5.3~V in Fig~\ref{fig:Gate_dep}a, may be explained by the
subsequent loss of $K'_\downarrow$ sub-bands at $B\simeq 8\,$T and $K'_\uparrow,K_\downarrow$ at $\simeq24\,$T in
the DOS. However, as the calculation neglects quantum interference effects in the Fabry-Perot
regime, such as the AB beating effect \cite{Cao}, we cannot expect to explain all features of the
measured magneto-conductance within our simple model.

In conclusion, we have investigated single walled carbon nanotubes up to very high magnetic fields.
The magneto-conductance of a quasi-metallic tube shows a peculiar double peak, which can be
explained in terms of spin split conduction bands, separated by a strong spin-orbit interaction,
which exceeds the Zeeman splitting. Our finding may open the path towards the application of CNTs
as highly efficient ballistic spin filters.

This research was funded by the Deutsche Forschungsgemeinschaft within GK 1570 and SFB 689 and
partially supported by EuroMagNET under the EU contract RII3-CT-2004-506239.


\begin{thebibliography}{}
\bibitem{SO} Therefore, $\Delta_{\text{SO}}$ is larger for our nanotube, compared to the value observed for the tube ($d \sim$ 5~nm) in Ref.\ \onlinecite{Kuemmeth}.
\end{thebibliography}
\end{document}